\documentclass[
    aps,
    reprint,
]{revtex4-2}

\usepackage{newtxtext,newtxmath}
\usepackage{bm}
\usepackage{physics}
\usepackage{xcolor}
\usepackage{mdframed}
\usepackage{comment}
\usepackage{verbatim}
\newcommand{\Ham}{\hat{\mathcal{H}}}

\newcommand{\cP}{\mathcal{P}}

\newcommand{\Sz}[1]{\hat S^{(#1)}_z}
\newcommand{\Sp}[1]{\hat S^{(#1)}_+}
\newcommand{\Sm}[1]{\hat S^{(#1)}_-}
\newcommand{\amp}[2]{\langle #1 \mid #2 \rangle}
\newcommand{\state}[1]{\ket{#1}}
\renewcommand{\eqref}[1]{Eq.~(\ref{#1})}
\newcommand{\secref}[1]{Sec.~\ref{#1}}
\usepackage[
colorlinks=true,
linkcolor=blue,
citecolor=magenta,
filecolor=cyan,
urlcolor=teal,
bookmarks=true,
bookmarksopen=true, %
breaklinks=true, %
]{hyperref}
\hypersetup{
    pdftitle={Spin Hamiltonian as Matrix-Free Linear Map},
    pdfauthor={Aditya Dev},
    pdfsubject={Matrix-free construction of generic spin Hamiltonians using mixed-radix product-basis indexing},
    pdfkeywords={spin Hamiltonians, exact diagonalization, mixed-radix indexing, Krylov methods, spin-boson systems, Jaynes-Cummings model},
    pdfcreator={LaTeX with hyperref},
    pdfproducer={pdfTeX}
}
\begin{document}

\title{Spin Hamiltonian as Matrix-Free Linear Map}

\author{Aditya Dev}
\affiliation{Department of Chemical \& Biological Physics, Weizmann Institute of Science, Rehovot 7610001, Israel}

\date{\today}

\begin{abstract}
We present an algorithm that computes the action of a generic spin Hamiltonian on a state vector on the fly, entirely avoiding explicit matrix assembly. This is achieved through mixed-radix indexing of the full tensor-product basis, which translates local spin operations into simple integer offsets. The result is an explicit framework for evaluating single- and two-site terms across arbitrary spin lattices, including mixed-spin systems. Our construction bridges the basis-indexing logic familiar from exact diagonalization with the matrix-free state-update philosophy of address-based frameworks. By writing the indexing logic in closed form, a single uniform loop applies to every site regardless of its local Hilbert-space dimension. The method is parallelizable and memory-conserving, and can be extended to restricted basis or truncated bosonic levels.
\end{abstract}

\maketitle
\section{Introduction}

\paragraph*{Background:} The exponential growth of Hilbert-space dimension is the main obstacle to the exact numerical simulation of quantum many-body systems.
For a lattice of $N$ spins, with spin $\{s_i\}_{i=1}^N$ at each site, the total dimension comes out to be:
\begin{equation}
    D = \prod_{i=1}^N d_i,
    \quad
    d_i = 2s_i + 1.
\end{equation}
Even when the Hamiltonian is sparse, assembling it numerically quickly becomes memory-limited.

The usual route to build the Hamiltonian $\Ham$ numerically is by repeated Kronecker products, or to use identities such as
\begin{equation}
    (A \otimes B)\,\mathrm{vec}(X) = \mathrm{vec}(BXA^{\mathsf T}),
\end{equation}
which avoid explicit matrix assembly but still require the matrix elements to be stored in some form~\cite{Henderson_Searle_1981,Roth_1934}, and usually involve repeated reshaping of the vector $X$.
For large systems, this assembly step can dominate the calculation. In practice, the simulation often never starts because the sparse matrix itself is already too big to hold in memory.

Modern Krylov methods, however, only require the map
\begin{equation}
    \state{y} = \Ham \state{x},
\end{equation}
not the explicit matrix $\Ham$. This fact is used by Lanczos- and Arnoldi-type solvers~\cite{lanczos1950iteration,ARPACK}.
If the action of $\Ham$ can be generated directly, the operator storage is removed completely, and the working memory drops
only to the vectors used by the algorithm.

The matrix-free approach to many-body simulation is, of course, not new in itself. In exact diagonalization, Nishimori and Taguchi already emphasized the usefulness of sparse-matrix techniques together with a bit representation for spin-$1/2$ Hamiltonians~\cite{NishimoriTaguchi1986,nishimori1992titpack}. Lin's exact-diagonalization work on quantum spin systems pushed the same philosophy to higher-spin models using integer-based encoding of many-body basis, where the central large-scale operation is repeated Hamiltonian action inside a Lanczos iteration rather than dense diagonalization~\cite{Lin1990}. Modern software packages continue this line of development. \texttt{QuSpin} provides symmetry-aware basis construction and exact dynamics for many-body systems, including homogeneous higher-spin and bosonic degrees of freedom, with operator construction through user-defined bases~\cite{QuSpinPartI2017,QuSpinPartII2019}. \texttt{XDiag}, in contrast, focuses on high-performance exact diagonalization for spin-$1/2$, fermionic, and $t$-$J$ systems, combining optimized basis encodings and symmetry reduction, defaulting to matrix-free operator application with explicit sparse construction available as an alternative~\cite{wietek2025xdiag,wietek2018sublattice}. Meanwhile, \texttt{dynamite} pushes the matrix-free point of view toward massively parallel workflows for large many-body spin-$1/2$ chains~\cite{KahanamokuMeyer2023dynamite,VanBeeumen2020matrixfree}. In these exact-diagonalization implementations, basis states are assigned integer labels (binary for spin-$1/2$ systems) and local operators are generated based on the explicitly provided basis or operator actions are simply translated into elementary updates of those labels; for spin-$1/2$ systems, this is especially natural because computational basis states are binary strings and local spin flips become bitwise operations. Once the local Hilbert-space dimension increases beyond two, that binary shorthand disappears; for homogeneous higher-spin systems, one may still use a fixed-base representation~\cite{Lin1990}, but genuinely mixed-spin systems require site-dependent dimensions.

A conceptually related development, closer in spirit to the on-the-fly generation of matrix elements, appears in the field of solid-state NMR simulations. Dumez \emph{et al.} observed that the time evolution of observables in large dipolar-coupled spin systems can be simulated with very high accuracy using low-order correlations in Liouville space. Rather than working with a full density-matrix vector, they maintain a sparse collection of coefficients in a reduced product-operator basis and directly update these coefficients using analytical formulas~\cite{Dumez2009CPL,Dumez2010JCP}. The underlying physical reason behind the analytical approach is that spin Hamiltonians generate rotations, so the action of a local propagator corresponds to a rotation in the space of expansion coefficients. Even though it requires explicit rotation-matrix construction, from an algorithmic viewpoint, this is a matrix-free update: local interaction rules generate coefficient-level changes rather than assembling the full Liouvillian as a global matrix and applying it via an explicit matrix–vector multiplication. Closely related ideas appear in the restricted-state-space methods developed by Kuprov and co-workers, where the Liouville-space basis is truncated in a spherical-tensor representation to retain only dynamically relevant spin-correlation and coherence orders~\cite{Kuprov2007SSR,Kuprov2008RSS,Krzystyniak2011DSS,Hogben2011Spinach}. As in the work of Dumez \emph{et al.}, the emphasis is on avoiding explicit matrix construction and propagating only the active components of the state vector, although here the reduction is achieved by restricting the state space itself. From a computational perspective, similar bookkeeping also appears in modern matrix-free packages such as \texttt{Rimu.jl}, where basis states (for bosonic or fermionic Fock configurations) are stored explicitly as address objects, and the operator is applied on-the-fly by generating connected states together with their coefficients, followed by a label-to-index lookup~\cite{vcufar2026rimu}. Although the underlying representations differ, the common theme remains: replace explicit matrix storage with local update rules acting on encoded basis states. Our approach complements these approaches.

In the present work, we make the indexing logic explicit and generic—covering arbitrary local spins (and bosonic Fock-states) with site-dependent Hilbert-space dimensions—and to pair it with an on-the-fly operator application that never forms an explicit matrix and in a form that plugs directly into matrix-free Krylov solvers. The main point is that for each target basis state, the source that contributes to it can be constructed algebraically, so the global Hamiltonian matrix never has to exist, even as a sparse object. The manuscript is organized as follows. We first motivate the target-driven matrix-vector viewpoint in \secref{sec:target_driven_mul} and then introduce the mixed-radix indexing of the full product basis in \secref{sec:mixed_radix_enumeration}. Next, we derive the one-site and two-site update rules that turn a generic spin Hamiltonian into an explicit matrix-free linear map, and benchmark it against conventional sparse assembly in \secref{sec:algorithm} and \secref{sec:benchmark}, respectively. In \secref{sec:generic_basis} we discuss how the same logic extends to generic restricted-basis settings. \secref{sec:limitation} before the conclusion, discusses the limitations of our method and how it compares to other techniques used in numerical simulation of quantum systems.

\section{Target-Driven Matrix-Vector Multiplication\label{sec:target_driven_mul}}
Explicit matrix assembly is naturally \emph{source driven}, i.e., one applies an operator on a state, finds all connected target states, and stores the resulting matrix elements. For a matrix-free execution, it is more convenient to reverse this logic, i.e.
we loop over the \emph{target} index of a basis state $F \in [0,D-1]$ and ask which source indices $I$ contribute to the amplitude~\cite{QuSpinPartI2017,QuSpinPartII2019,DanceQ_2025,KahanamokuMeyer2023dynamite,VanBeeumen2020matrixfree}
\begin{equation}
    y_F = \amp{F}{y} = \bra{F}\Ham\ket{x}.
\end{equation}
This target-driven form is well-suited for parallel execution because each thread writes only to its own component $y_F$.
No synchronization is required as long as the source vector $x_I$ is read only. The trade-off is that the source-vector reads $x_I$ are scattered, leading to a random-access pattern with limited cache locality. This is the same as in a sparse matvec, and partly explains why our matrix-free and sparse matvec reach comparable steady-state matvec rates in our benchmarks (\secref{sec:benchmark}).
\begin{mdframed}[backgroundcolor=gray!5, linecolor=gray!30, roundcorner=3pt]
\textbf{Notation guide.}\label{notation_guide}
\begin{itemize}
    \item $\state{F}$ and $\state{I}$ denote target and source basis states.
    \item $x_I = \amp{I}{x}$ and $y_F = \amp{F}{y}$ are vector components.
    \item Site labels are written as superscripts in parentheses, e.g. $\Sz{i}$.
    \item Local spin projections are $m_i \in \{-s_i,\dots,s_i\}$.
    \item Shifted local indices are $n_i = m_i + s_i \in \{0,\dots,d_i-1\}$.
\end{itemize}
\end{mdframed}
For spin-$1/2$ systems, this same idea appears as bit-update logic~\cite{QuSpinPartI2017,QuSpinPartII2019,KahanamokuMeyer2023dynamite,VanBeeumen2020matrixfree}. The present formulation maintains the same computational structure while replacing binary arithmetic with mixed-radix arithmetic, enabling arbitrary local dimensions to be handled uniformly. The key point is that the change induced by local operators affects only a few local quantum numbers, thereby shifting the basis deterministically.
The entire method, therefore, reduces to two tasks:
\begin{enumerate}
    \item map basis states $\state{n_1,\dots,n_N}$ to integer indices and back;
    \item express each local operator as a small deterministic index shift.
\end{enumerate}

\section{Mixed-Radix Indexing of the Product Basis\label{sec:mixed_radix_enumeration}}
We enumerate the product basis in lexicographic order~\cite{aigner2007course}. We define a multiplier
\begin{equation}
    M_k = \prod_{\ell=k+1}^{N} d_\ell,
    \quad
    M_N = 1.
\end{equation}
Then we define a map that associates a basis state \(
\state{n_1,n_2,\dots,n_N}
\) to a global index, i.e.
\begin{equation}
   \state{n_1,n_2,\dots,n_N} \to I = \sum_{k=1}^{N} n_k M_k.
    \label{eq:encode}
\end{equation}
\eqref{eq:encode} is the mixed-radix representation of the tuple $(n_1,\dots,n_N)$~\cite{Knuth1970TheAO,aigner2007course} and exactly the rule underlying tensor flattening packages when axis lengths are site -dependent~\cite{kolda2009tensor,Ansel_PyTorch_2_Faster_2024,numpy}.  It is useful to separate the three levels of generality here.

\emph{First}, if every site carries spin-$1/2$, then every local dimension is $d_k=2$ and multipliers $M_k$ are powers of $2$. The global
state labels become binary numbers. This is the
setting in which bit masks, XORs, and shifts are perfect candidates~\cite{QuSpinPartI2017,KahanamokuMeyer2023dynamite,wietek2025xdiag}.

\emph{Second}, if every site carries the same higher spin $s$, then every local dimension is still the same, but now
\begin{equation}
    d_k = d = 2s+1
\end{equation}
for all $k$. In this case, we have a uniform base-$d$ representation:
\begin{equation}
    I = \sum_{k=1}^N n_k d^{N-k}.
\end{equation}
This is already enough to cover, for example, spin-$1$ chains and spin-$3/2$ chains, and it is the natural setting in which the higher-spin exact-diagonalization literature begins to move beyond binary encodings~\cite{Lin1990,QuSpinPartII2019}.

\emph{Third}, for a fully generic system, the local spins may differ from site to site, and then the
representation becomes mixed-radix, with the global index encoded through the ``generic" place value system given in \eqref{eq:encode}, where the multipliers $\{M_k\}$ are evaluated for the site-dependent dimensions $\{d_\ell\}$.
The conceptual gain of writing the method this way is that both the binary and the uniform higher-spin case appear simply as a special case of this indexing rule.

The local state index $n_k$
is extracted by division by the corresponding multiplier $M_k$ followed by reduction modulo $d_k$,
i.e., the inverse map is
\begin{equation}
    n_k = \left\lfloor \frac{I}{M_k} \right\rfloor \bmod d_k,
    \label{eq:decode}
\end{equation}
and the corresponding magnetic projection follows from the notation guide of~\secref{sec:target_driven_mul} as
\begin{equation}
    m_k = n_k - s_k.
\end{equation}
This decoding step is the key to the on-the-fly algorithm. Once the local digits $\{n_k\}$ are known, any local operator can be handled without ever assembling the Hamiltonian.
\section{Algorithm\label{sec:algorithm}}
Here we present the core algorithmic recipe. For simplicity of presentation, we separate the single-site and two-site terms into two different subsections. Readers who prefer an executable recipe over several pages of algebra may find it helpful to go through the pseudocode in Appendix~\ref{appen:pseudocode} before reading the derivation below. The appendix provides a compact implementation-oriented summary that mirrors the derivation presented here.
\subsection{Single-Site Terms}
Consider a generic one-body Hamiltonian
\begin{equation}
    \Ham_1
    = \sum_{i=1}^{N}
    \left(
        h^{(i)}_z \Sz{i}
        + h^{(i)}_+ \Sp{i}
        + h^{(i)}_- \Sm{i}
    \right),
\end{equation}
where hermiticity requires $h^{(i)}_- = (h^{(i)}_+)^{*}$, with the ladder couplings related to physical transverse fields through $h^{(i)}_+ = (h^{(i)}_x - i h^{(i)}_y)/2$. The split into ladder pieces is purely algorithmic, since each ladder action shifts a single local digit by $\pm 1$ \footnote{This approach also extends to systems described by Lindblad generators, although the coefficients are not generally related as simply as in the Hermitian case.}.
For a fixed target index $F$, the local target digit $n_i$ at site $i$ is obtained using \eqref{eq:decode}. Once $n_i$ is known, the diagonal term is:
\begin{equation}
    y_F \mathrel{+}= h^{(i)}_z\, m_i \, x_F.
\end{equation}
For the raising operator, we have
\begin{equation}
    \hat S_+ \state{m}
    =
    \sqrt{s(s+1)-m(m+1)} \, \state{m+1}.
\end{equation}
We repeat the same $n_i$ extraction using \eqref{eq:decode}. If that target digit is $n_i$, then a contribution from $\Sp{i}$ can only originate from the source digit $n_i-1$, since the operator must raise the source projection by one unit in order to reach the target. Consequently:
\begin{enumerate}
    \item if $n_i = 0$, there is no source state and the contribution is zero;
    \item otherwise the source index is $I = F - M_i$;
    \item the source magnetic quantum number is $m_i^{(I)} = (n_i - s_i) - 1$;
    \item the update is
    \begin{equation}
        y_F \mathrel{+}=
        h^{(i)}_+
        \sqrt{s_i(s_i+1)-m_i^{(I)}(m_i^{(I)}+1)}
        \, x_I.
    \end{equation}
\end{enumerate}

The lowering term is analogous, with a negative index shift.
\begin{equation}
    \hat S_- \state{m}
    =
    \sqrt{s(s+1)-m(m-1)} \, \state{m-1},
\end{equation}
a target digit $n_i$ can only be reached from the source digit $n_i+1$. The source must therefore
lie one unit \emph{above} the target on site $i$. This gives:
\begin{enumerate}
    \item if $n_i = d_i-1$, there is no admissible source state, because the target already sits at
    the maximal local digit;
    \item otherwise the source index is $I = F + M_i$, since increasing the source digit by one
    raises the global mixed-radix index by exactly one multiplier $M_i$;
    \item the source magnetic quantum number is $m_i^{(I)} = (n_i-s_i)+1$;
    \item the contribution takes the form
\begin{equation}
    y_F \mathrel{+}=
    h^{(i)}_-
    \sqrt{s_i(s_i+1)-m_i^{(I)}(m_i^{(I)}-1)}
    \, x_I.
\end{equation}
\end{enumerate}
Thus, the one-site off-diagonal structure is already fully determined by the sign of the ladder move: $\Sp{i}$ shifts the source index to $F-M_i$, whereas $\Sm{i}$ shifts it to $F+M_i$. The coefficient in each case is the standard local ladder matrix element evaluated on the reconstructed source state.

\subsection{Two-Site Terms\label{sec:two-site-terms}}
For pair interactions, it is useful to write the Hamiltonian as
\begin{equation}
    \Ham_2
    =
    \sum_{(i,j)\in\cP}
    \sum_{\alpha,\beta \in \{z,+,-\}}
    J_{ij}^{\alpha\beta}\,
    \hat S_{\alpha}^{(i)} \hat S_{\beta}^{(j)},
    \label{eq:pairham}
\end{equation}
where $\cP$ is the set of active interacting pairs and the couplings $J_{ij}^{\alpha\beta}$ are assumed to be known and precomputed. Hermiticity imposes the relations $(J_{ij}^{\alpha\beta})^{*} = J_{ij}^{\bar{\alpha}\bar{\beta}}$, where $\bar{z}=z$ and $\overline{\pm}=\mp$. This covers Ising terms, exchange terms, anisotropic couplings, and cross-terms such as $\hat S_x^{(i)} \hat S_y^{(j)}$ after expansion
into standard spin ladder operators.

The two-site update follows the same logic as the one-body case, except that two local digits must change simultaneously. Each local operator carries a shift
\begin{equation}
    \delta(z)=0,
    \quad
    \delta(+)=+1,
    \quad
    \delta(-)=-1.
\end{equation}
For a fixed target index $F$ and a term
\(
\hat S_{\alpha}^{(i)} \hat S_{\beta}^{(j)}
\),
the candidate source digits are
\begin{equation}
\label{eq:src_digit_two_site}
    n_i^{(\mathrm{src})} = n_i - \delta(\alpha),
    \qquad
    n_j^{(\mathrm{src})} = n_j - \delta(\beta),
\end{equation}
with all other digits unchanged. The corresponding source index is therefore
\begin{equation}
    I = F - \delta(\alpha) M_i - \delta(\beta) M_j.
    \label{eq:sourceoffset}
\end{equation}
This source index is valid only if the implied source digits lie in the allowed local ranges, i.e.,
\begin{equation}
    0 \le n_i^{(\mathrm{src})} \le d_i-1,
    \quad
    0 \le n_j^{(\mathrm{src})} \le d_j-1.
\end{equation}

\paragraph*{Algorithmic kernel.}
For each target amplitude $y_F$, the pair contribution is evaluated by a short and reusable sequence:
\begin{enumerate}
    \item decode the target labels $n_i$ and $n_j$ using $F$, hence the target projections $m_i=n_i-s_i$ and $m_j=n_j-s_j$;
    \item for each active pair $(i,j)$ and each operator channel $(\alpha,\beta)$, construct the candidate source digits $n_i^{(\mathrm{src})}$ and $n_j^{(\mathrm{src})}$ using \eqref{eq:src_digit_two_site};
    \item reject the channel if either source digit falls outside the allowed local range;
    \item otherwise reconstruct the source index $I$ using \eqref{eq:sourceoffset};
    \item evaluate the two local matrix elements on the reconstructed source state and accumulate the contribution into $y_F$.
\end{enumerate}
The diagonal pair channel $(\alpha,\beta)=(z,z)$ is handled by the same kernel without special-casing: both shifts vanish, the source digit checks are trivially satisfied, the source index reduces to $I=F$, and the contribution collapses to $J_{ij}^{zz}\,m_i m_j\,x_F$.
If the source is admissible, the update may be written compactly as
\begin{equation}
    y_F \mathrel{+}=
    J_{ij}^{\alpha\beta}\,
    \bra{m_i^{(F)}}\hat S_{\alpha}^{(i)}\ket{m_i^{(I)}}
    \bra{m_j^{(F)}}\hat S_{\beta}^{(j)}\ket{m_j^{(I)}}
    \, x_I,
\end{equation}
where $m_i^{(I)}=n_i^{(\mathrm{src})}-s_i$ and $m_j^{(I)}=n_j^{(\mathrm{src})}-s_j$. The two-site terms are therefore reduced to two admissibility checks, one integer offset, and two local coefficients.

\paragraph*{Connection with standard basis.}
To get an intuitive sense of the source-target index relationship that we have presented in previous sections, let's rewrite everything explicitly in the product-basis language. If the target state corresponding to the index $F$ is
\begin{equation}
\ket{F}\equiv\ket{m_1,\dots,m_i,\dots,m_N},
\end{equation}
then the matrix element
\begin{equation}
\bra{F}\hat S_{\alpha}^{(i)}\ket{I}
\end{equation}
can be nonzero only if the source state $\ket{I}$ matches $\ket{F}$ on all sites $k \neq i$. The operator acts nontrivially only on site $i$, so any difference between $\ket{I}$ and $\ket{F}$ is confined to that particular site $(i)$ and is fixed by the ladder shift associated with $\alpha$. More precisely, the source state must have the form
\begin{equation}
    \ket{I}
    =
    \ket{m_1,\dots,m_i-\delta(\alpha),\dots,m_N},
\end{equation}
provided that the shifted local projections still lie within the allowed ranges.

In the mixed-radix digit representation, this becomes
\begin{align}
    n_k^{(I)} = n_k^{(F)} \quad (k\neq i), \;\& \;
    n_i^{(I)} = n_i^{(F)}-\delta(\alpha).
\end{align}
Since only the $i$th digit changes, the global source index differs from the target index only through the corresponding place value. Changing the digit at site $i$ by one shifts the total integer index by exactly $M_i$, so
\begin{equation}
    I = F - M_i.
\end{equation}
For a genuine two-site term, the two place-value shifts simply add. Thus, if
\begin{equation}
    \ket{F}=\ket{m_1,\dots,m_i,\dots,m_j,\dots,m_N},
\end{equation}
then the channel $\Sp{i}\Sm{j}$ gives
\begin{align}
    I &= F - M_i + M_j, \notag\\
    y_F \mathrel{+}={}&
    J_{ij}^{+-}
    \sqrt{s_i(s_i+1)-m_i^{(I)}(m_i^{(I)}+1)}
    \notag\\
    &\times
    \sqrt{s_j(s_j+1)-m_j^{(I)}(m_j^{(I)}-1)}
    \, x_I,
\end{align}
and the remaining ladder combinations follow by the corresponding sign choices in \eqref{eq:sourceoffset}. The diagonal contribution does not change the index, so
\begin{equation}
    \Sz{i}\Sz{j}: \quad
    I = F, \qquad
    y_F \mathrel{+}= J_{ij}^{zz}\, m_i m_j\, x_F.
\end{equation}
\eqref{eq:sourceoffset} removes the need for bespoke global indexing logic for each Hamiltonian term. A two-site contribution is fully characterized by three objects: the pair $(i,j)$, the local shifts $\delta(\alpha)$ and $\delta(\beta)$, and the coefficient table $J_{ij}^{\alpha\beta}$. The remainder of the computation is reduced to integer arithmetic.

\paragraph*{Higher local powers and spin-boson systems.} The construction is not limited to single powers of spin operators. Higher-order terms, such as zero-field splitting, involve local transitions that shift the digit by multiple units. Similarly, truncated bosonic systems can also be treated assuming creation and annihilation operators correspond to unit shift operators with appropriate square-root coefficients. We provide the details and explicit examples for higher power spin terms and spin-boson (Jaynes--Cummings) models in Appendix~\ref{appen:higher_powers} and Appendix~\ref{appen:spin_boson_jc}, respectively.

\section{Benchmark against Explicit Sparse Assembly\label{sec:benchmark}}
The cleanest numerical comparison remains the standard exact-diagonalization workflow. We therefore benchmark the present matrix-free kernel against explicit sparse assembly across three testbeds~\cite{Dev2026MatrixFreeCode}. The benchmark parameters are:
\begin{enumerate}
    \item \textbf{Uniform spin-$1$ PBC (Periodic boundary condition) chain:} $N=15$, nearest-neighbor XXZ with $J_{xy}=1.0$, $J_z=1.0$, and staggered field $h_i=\pm 0.2$ (alternating by site), giving $D=3^{15}=14{,}348{,}907$.
    \item \textbf{Mixed-spin PBC chain:} local pattern $(1/2)^6(1)^6(3/2)^6$, again with nearest-neighbor XXZ couplings $J_{xy}=J_z=1.0$ and staggered field $h_i=\pm0.2$, giving $D=2^6 3^6 4^6 = 24^6 = 191{,}102{,}976$.
    \item \textbf{QuSpin example-12 geometry (2D):} spin-$1/2$ lattice $5\times 5$ ($D=2^{25}=33{,}554{,}432$), with static nearest-neighbor XXZ couplings $J_1=1.0$ and driven diagonal/anti-diagonal XXZ couplings $J_2\cos(\Omega t)$ at $t=0$, where $J_2=0.5$ and $\Omega=8.0$. Threading in this run was $\texttt{OMP}=60$, $\texttt{MKL}=1$, and Numba threads $=60$.
\end{enumerate}

\begin{table*}[!ht]
\caption{Benchmark results for explicit sparse assembly versus matrix-free action. Timings are in seconds; storage is operator-storage cost in MB.}
\begin{ruledtabular}
\scriptsize
\begin{tabular*}{\textwidth}{@{\extracolsep{\fill}}lcccccccc}
Case & Bkd & Setup & 1st apply & Eigensolve & Total & Matvec/s & Storage (MB) & Residual \\
\hline
C1 & Sparse & $5.454\times10^{1}$ & $2.970\times10^{-1}$ & $3.849\times10^{1}$ & $9.333\times10^{1}$ & $3.281$ & $2.406\times10^{3}$ & $1.675\times10^{-10}$ \\
C1 & MF & $2.123$ & $1.610\times10^{-1}$ & $2.928\times10^{1}$ & $3.157\times10^{1}$ & $6.137$ & $4.578\times10^{-4}$ & $2.961\times10^{-10}$ \\
\hline
C2 & Sparse & $1.284\times10^{3}$ & $8.403$ & $1.332\times10^{3}$ & $2.624\times10^{3}$ & $1.139\times10^{-1}$ & $4.807\times10^{4}$ & $2.325\times10^{-9}$ \\
C2 & MF & $2.500\times10^{1}$ & $2.179\times10^{1}$ & $2.748\times10^{3}$ & $2.795\times10^{3}$ & $4.387\times10^{-2}$ & $8.240\times10^{-4}$ & $7.296\times10^{-10}$ \\
\hline
C3 & Sparse & $2.745\times10^{3}$ & $1.044$ & $1.805\times10^{3}$ & $4.551\times10^{3}$ & $8.612\times10^{-1}$ & $3.277\times10^{4}$ & $7.749\times10^{-11}$ \\
C3 & MF & $3.884$ & $1.225$ & $8.252\times10^{2}$ & $8.303\times10^{2}$ & $7.642\times10^{-1}$ & $2.670\times10^{-3}$ & $1.353\times10^{-10}$ \\
\hline
\multicolumn{9}{l}{\footnotesize C1: spin-$1$ chain ($N=15$). C2: mixed-spin chain $(1/2)^6(1)^6(3/2)^6$. C3: 2D spin-$1/2$ $5\times5$ (QuSpin example-12 model).}\\
\multicolumn{9}{l}{\footnotesize Bkd abbreviations: Sparse = explicit sparse assembly; MF = matrix-free}\\
\end{tabular*}
\end{ruledtabular}
\label{tab:benchmark}
\end{table*}

Across all three cases, the ground-state energies agree at near machine precision ($|\Delta E_0|\sim10^{-14}$). The strongest and most consistent advantage of the present approach is operator storage: the matrix-free metadata remain in the KB range, while explicit sparse storage ranges from GB to tens of GB (storage ratios from $\sim 10^6$ to $\sim 10^8$). This is the core practical novelty of the mixed-radix target-driven map: it keeps the update local and exact while avoiding global matrix assembly.

To make the scaling trend explicit, we summarize each testbed separately; Table~\ref{tab:benchmark} presents the findings. In the uniform spin-$1$ chain ($D=14{,}348{,}907$), matrix-free improves total wall time by about a factor of three ($93.3\,\mathrm{s}\to31.6\,\mathrm{s}$), with both faster first apply and faster eigensolve in this run, while reducing operator storage by about $5.26\times10^6$. In the mixed-spin chain ($D=191{,}102{,}976$), matrix-free still provides a dramatic storage reduction ($\sim5.83\times10^7$), and setup time drops by almost two orders of magnitude; however, sparse retains a slight total-time advantage because its first-apply and eigensolve phases are faster once the matrix is already assembled. In the $5\times5$ 2D example-12 geometry ($D=33{,}554{,}432$), matrix-free is favorable in both setup and total time (about $5.5\times$ lower total time) while again reducing storage by more than seven orders of magnitude ($\sim1.23\times10^7$). These three cases show the practicalities of the presented algorithm. The matrix-free strategy is strongest when explicit assembly and operator storage dominate the workflow, especially for large parameter sweeps or repeatedly rebuilt Hamiltonians. Its weak point is that the steady-state sparse matvec and eigensolve can still be faster after assembly is already paid.
\subsection{Complexity and memory scaling}
Each matvec performs a constant amount of work per target index $F$: a digit decode costs $O(1)$ per site for fixed-width integers, hence $O(N)$ for all sites, an $O(N)$ sweep over single-site terms, and an $O(|\cP|)$ sweep over pairs (with up to nine $(\alpha,\beta)$ channels per pair). The total time complexity is therefore
\begin{equation}
    T_{\mathrm{matvec}} = O\!\left(D\,(N + |\cP|)\right),
\end{equation}
which has the same asymptotic shape as a CSR sparse matvec, where the cost scales as $O(D \cdot k)$ with $k$ the average number of nonzeros per row. The two algorithms thus share the same $D$-leading scaling, but the constants differ in two important respects. First, the operator-side memory drops to
\begin{equation}
    M_{\mathrm{op}} = O(N + |\cP|),
\end{equation}
because the matrix-free kernel only stores $\{d_i\}$, $\{M_i\}$, the pair list, and the coupling tables, never the row/column structure. Second, no assembly passes over $D$ is required, removing a one-time Hamiltonian assembly cost. The cost of evaluating local ladder matrix elements is absorbed into the constants and does not affect the scaling.
\section{Generic Restricted-Basis Formalism\label{sec:generic_basis}}
The matrix-free construction removes the cost of storing $\Ham$, leaving the dense vector itself as the primary memory bottleneck. However, when vector storage becomes the limiting factor, the idea presented can be extended to a restricted basis as well.

In the physics literature, there are two conceptually different ways to reduce this ``state-storage cost''. The first and most straightforward one is basis restriction: (e.g., a fixed magnetization sector or local truncation), which shrinks the active state space itself; one keeps only the admissible states, and the vector retains fewer physical entries. This idea has been extensively used in magnetic-resonance spin dynamics, where restricted spherical-tensor or product-operator bases make the simulation of large spin systems possible by retaining only the important spin-correlation components~\cite{Kuprov2007SSR,Kuprov2008RSS,Krzystyniak2011DSS,Hogben2011Spinach,Dumez2009CPL,Dumez2010JCP}. Second, compressed representations like the tensor-train (TT) or matrix-product-state (MPS) formats keep the large physical space and store its amplitudes in a factorized form rather than as a single flat array. They remain unsuitable for time-domain simulations and irregular spin systems and would require a different implementation language, which falls outside the scope of the present manuscript. We therefore restrict our attention to the restricted basis set.
\subsection{Restricted basis storage}
Let the retained basis states be labeled by strings $\boldsymbol{\lambda}=(\lambda_1,\lambda_2,\dots,\lambda_N)$, where each $\lambda_i$ is some local descriptor or a quantum number. The storage problem then splits into two different cases.

\paragraph*{Case 1 (direct-product retained basis).} If the retained basis is still a direct product of local truncated sets, for example, a bosonic mode truncated to $d_i^{(\mathrm{res})}$ retained levels at site $i$ or a spin site with a reduced local set, so that $\lambda_i \in \{0,\dots,d_i^{(\mathrm{res})}-1\}$ independently at each site, then the global storage index is again a mixed-radix integer:
\begin{equation}
    I_{\mathrm{res}}(\boldsymbol{\lambda}) = \sum_{k=1}^{N} \lambda_k M_k^{(\mathrm{res})},
    \qquad
    M_k^{(\mathrm{res})} = \prod_{\ell=k+1}^{N} d_\ell^{(\mathrm{res})}.
\end{equation}
This is exactly the same as the full-basis indexing but applied to reduced local dimensions. In this case, the storage index of a retained state is obtained directly from its local labels by restricted mixed-radix arithmetic.

\paragraph*{Case 2 (globally constrained sector).} By contrast, if there are restrictions imposed by some global constraint, the bookkeeping becomes more involved. A useful example is a fixed-total-magnetization sector, $M = \sum_i m_i$. Each retained state is still a product state, but the admissible tuples are scattered non-contiguously through the full Hilbert space, so one can no longer assign storage indices by a simple mixed-radix rule over the retained basis. In that case, one typically stores the admissible tuples  $\boldsymbol{\lambda}$ explicitly, for example, in a packed array, together with a lookup structure over the retained basis labels; i.e., a function that checks the validity of that state under the global constraint/symmetry and returns its position index in the array. This bookkeeping is not new: for example, \texttt{Rimu.jl} implements it using a hash-table dictionary with a fast hash function for basis labels~\cite{vcufar2026rimu}. Hashing is not the only choice for indexing scattered basis labels; one can instead use Lehmer-code-based lexicographic indexing or some kind of combinatorial ranking, where the packed index is obtained by counting the number of admissible configurations that precede the state in the chosen ordering. This combinatorial idea goes back to Lehmer's work ~\cite{lehmer1960teaching}, and has been implemented in exact diagonalization numerical packages~\texttt{DanceQ}~\cite{DanceQ_2025}. An explicit example implementation of the lookup approach has been given in Appendix~\ref{sec:lookup_mech}
\subsection{Source reconstruction and membership test}
Suppose a Hamiltonian term acts on site $i$, or on a pair $(i,j)$. For a retained target state, the inverse local move produces the candidate source labels $\lambda_i^{(\mathrm{src})}$ (and $\lambda_j^{(\mathrm{src})}$ for two-site interactions), with all inactive labels unchanged. For example, in a fixed-$M$ sector, an $\Sp{i}\Sm{j}$ term acting on a target state requires the source state to have one \emph{less} quantum of spin at site $i$ and one \emph{more} at site $j$. The bookkeeping question is then how to convert those candidate source labels into the stored index. Throughout this subsection, $F_{\mathrm{tgt}}$ and $I_{\mathrm{src}}$ denote dense packed-storage indices in the retained basis.
\paragraph*{Case 1 (direct-product retained basis).} In the direct-product case, the stored source index is recovered by the same offset rule as in the full basis, i.e.,
  \begin{equation}
      I_{\mathrm{src}} = F_{\mathrm{tgt}} + \Delta\lambda_i M_i^{(\mathrm{res})} + \Delta\lambda_j M_j^{(\mathrm{res})}.
  \end{equation}
  where
  \begin{equation}
      \Delta\lambda_i = \lambda_i^{(\mathrm{src})} - \lambda_i^{(\mathrm{tgt})}, \quad \Delta\lambda_j = \lambda_j^{(\mathrm{src})} - \lambda_j^{(\mathrm{tgt})},
  \end{equation}
    Here $I_{\mathrm{src}}$ is obtained directly by restricted mixed-radix arithmetic.

\paragraph*{Case 2 (globally constrained sector).} In the globally constrained case, by contrast, the candidate source labels must be looked up in the retained basis:
  \begin{equation}
      I_{\mathrm{src}} = \mathrm{lookup}\!\left(\boldsymbol{\lambda}^{(\mathrm{src})}\right).
  \end{equation}
The update is accepted only if the lookup succeeds, i.e., if $\boldsymbol{\lambda}^{(\mathrm{src})}$ lies within the restricted basis. Equivalently, $\mathrm{lookup}(\cdot)$ returns a packed index on success and a failure flag (or sentinel) otherwise. This is crucial: the \emph{local physics} (undoing the ladder operator) is the same as in the full-basis case, but the \emph{global bookkeeping} changes.

The main computational trade-off in case 2 is the overhead of restricted-basis membership tests. In the direct-product setting (case 1), the packed source index is recovered by a direct constant-time integer offset calculation. In the globally constrained setting (case 2), one must instead perform a membership test via $\mathrm{lookup}(\boldsymbol{\lambda}^{(\mathrm{src})})$ for every candidate source state. This may introduce overhead, but that cost is the trade-off for a smaller propagated vector. If the retained tuples are simply stored in sorted order, a lookup can be implemented with binary search, yielding $O(\log N_{\mathrm{res}})$ complexity per candidate~\cite{DanceQ_2025}. For better performance in static restricted bases, one can instead precompute a minimal perfect hash function (MPHF) that maps the scattered physical identifiers $\{\lambda^1, \lambda^2, \ldots\}$ to $\{0, 1, 2, \ldots\}$ with deterministic $O(1)$ lookup~\cite{vcufar2026rimu}. Alternatively, if the constraint is structured enough, the lookup can be done using a Lehmer-code-style construction, i.e., the state itself carries enough combinatorial information to determine its packed position. The latter avoids a large global hash table while still returning the same index needed by the matrix-free update~\cite{DanceQ_2025}.

A final implementation remark concerns the index range. In a fixed-width realization that stores the global basis index in an unsigned $w$-bit machine integer, exact single-word indexing requires
\begin{equation}
      D=\prod_{i=1}^N d_i < 2^w.
\end{equation}
On standard 64-bit hardware, this means $D<2^{64}$ for a one-word linear index. This bound is realistic for low-level implementations that rely on native machine integers, but it is not a
fundamental limit on what a computer can represent. One may certainly store larger integers using multiword or arbitrary-precision arithmetic. The price is that index decoding, offset updates,
and array access are then no longer native constant-cost machine-word operations. More importantly, a flat state vector of length $D$ becomes impossible to store long before that formal bound is reached. The quantity $2^w$ can therefore be best interpreted as the ceiling for a \emph{single-word linear-index implementation}, not as an absolute ceiling on computable Hilbert-space dimension.
\section{Scope, Limitations, and Use Cases\label{sec:limitation}}
While our formalisms bypass the exponential scaling of memory for operators, full-basis propagation still scales with the state dimension, which is exponential too. Restricted bases can delay this limit when physically justified. A second limitation is dynamical: the method is fundamentally constrained by the physics of correlations~\footnote{The author deliberately avoids using the word entanglement — despite enormous social pressure from the field.}. If the system is chaotic or if there's a rapid correlation growth, even restricted bases can expand rapidly. No dynamic pruning threshold can prevent the basis from expanding to encompass the entire Hilbert space~\cite{fast_scramblers, PhysRevLett.70.3339}. In these regimes, dynamic basis restriction merely delays the eventual return to full Hilbert-space scaling. This limitation is intrinsic to exact numerical simulation rather than to the indexing scheme itself.
\paragraph*{Comparison with state-of-the-art techniques.}
It is useful to position the present method against the standard families used in the same regime, because each excels in a distinct part of the problem space, and the matrix-free mixed-radix approach is advantageous only in a well-defined slice of it. Sparse exact diagonalization with sparse matvec uses tightly optimized kernels and is usually the fastest steady-state option for repeated applications at fixed parameters; the present kernel matches that asymptotic time complexity but pays no assembly
cost and stores only $O(N+|\cP|)$ operator metadata, so the crossover favors matrix-free whenever the Hamiltonian is rebuilt frequently, or sparse storage simply does not fit in memory. Table~\ref{tab:benchmark} illustrates both behaviors: sparse retains a small steady-state advantage in case~C2, yet operator-memory ratios of $10^6$--$10^8$ make matrix-free the only practical option as $D$ grows. Symmetry-aware packages such as \texttt{QuSpin}, \texttt{XDiag}, and \texttt{dynamite} reduce the Hilbert-space dimension by orders of magnitude when strong global symmetries are present, and they dominate any flat-basis approach in those sectors; the present method does not exploit such symmetries. Tensor-network methods compress the wavefunction itself and operate at sub-exponential cost whenever entanglement is bounded, but they necessarily introduce truncations that become uncontrolled for highly excited states with rapid correlation spreading; the present flat-basis propagation remains exact regardless.

Bringing these comparisons together, the matrix-free mixed-radix construction is a stronger tool when several conditions coincide: (i) the system has site-dependent dimensions or mixed spin and bosonic content, so binary or uniform-base shortcuts do not apply; (ii) global symmetries do not produce a useful block reduction, or are explicitly broken by the model under study; (iii) the Hamiltonian must be repeatedly rebuilt, as in parameter sweeps, time-dependent drives, or disorder averaging; and (iv) exact propagation is required, while $D$ is still small enough for the dense vector to fit. In these regimes, neither sparse assembly, nor symmetry sectors, nor tensor-network truncation, provides what the present construction does: a single uniform on-the-fly kernel that works at machine precision, with operator memory independent of $D$.
\section{Conclusion}
We have written the action of a generic spin Hamiltonian as a matrix-free linear map by expressing the product basis as a mixed-radix integer system. In the full product basis, one-site and two-site spin operators then reduce to deterministic source-index reconstructions plus local matrix-element evaluations. The result is a direct target-driven matvec that removes explicit sparse-matrix assembly and is naturally compatible with Krylov methods. We also extend the algorithmic core to higher-order spin terms and to spin-bosonic mixed systems.

We benchmark our method against explicit sparse matrix-vector multiplication and highlight the trade-off: explicit sparse assembly can yield faster repeated matvecs once the matrix is built, but the matrix-free route is better when operator assembly is itself the bottleneck. This makes the present construction particularly well-suited to workflows involving parameter sweeps, time-dependent or dynamically modified couplings, and mixed-spin geometries, where repeated assembly would be costly or prohibitive.

Apart from the full product basis, the same logic extends naturally to a restricted basis. Once the operator-memory bottleneck is removed by the matrix-free update, the dense-vector storage becomes the primary constraint, and basis restriction offers a direct way to reduce the propagated state space. In that setting, direct-product truncations preserve the same mixed-radix offset arithmetic, whereas globally constrained sectors replace offsets with a membership test via a label-to-index lookup. The mixed-radix matrix-free construction is thus a useful framework for many-body linear maps in reduced state spaces as well.
\section*{Acknowledgments}
The author is grateful to Prof.~Ilya Kuprov for his insightful comments and suggestions on the manuscript, and for drawing attention to the work of Dumez \emph{et al.}, which significantly improved the presentation. The author thanks Prof.~Vishvendra Singh Poonia for valuable feedback and for the opportunity to work in his group as a research fellow, where the initial idea for this work was conceived. Helpful discussions with non-human colleagues are also gratefully acknowledged; large language models were used for editorial suggestions and for assistance in drafting the Julia and Python scripts.
\section*{Code Availability}
The scripts used to generate Table~\ref{tab:benchmark} are available online~\cite{Dev2026MatrixFreeCode}.
\bibliography{reference}
\pagebreak
\appendix
\onecolumngrid
\section{Matrix-Free Matvec Pseudocode\label{appen:pseudocode}}
The preceding formulas can be summarized by introducing two local functions. The first is the index
shift
\begin{equation}
    \delta(z)=0,
    \quad
    \delta(+)=+1,
    \quad
    \delta(-)=-1,
\end{equation}
and the second is the local matrix element
\begin{equation}
    \Lambda(\alpha;s,m)=
    \begin{cases}
        m, & \alpha=z, \\
        \sqrt{s(s+1)-m(m+1)}, & \alpha=+, \\
        \sqrt{s(s+1)-m(m-1)}, & \alpha=-.
    \end{cases}
\end{equation}
With this notation, diagonal and off-diagonal terms are handled by the same source-reconstruction
rule.

The required inputs are:
\begin{enumerate}
    \item the number of sites $N$;
    \item the local spins $\{s_i\}_{i=1}^N$, local dimensions $d_i=2s_i+1$, and multipliers $M_i$;
    \item the input vector $\state{x}\in\mathbb{C}^D$ and output vector $\state{y}\in\mathbb{C}^D$;
    \item the one-site couplings $h_z^{(i)}, h_+^{(i)}, h_-^{(i)}$;
    \item the active pair list $\cP$;
    \item for each $(i,j)\in\cP$, the coefficient table $J_{ij}^{\alpha\beta}$ with
    $\alpha,\beta\in\{z,+,-\}$.
\end{enumerate}
\begin{widetext}
\begin{verbatim}
function MATVEC!(y, x, spins, dims, mults, onsite, pair_terms)
    # spins[i]   = s_i
    # dims[i]    = d_i = 2 s_i + 1
    # mults[i]   = M_i
    # onsite     = {hz, hp, hm}
    # pair_terms = list of pairs (i, j, J) with J[alpha, beta]
    #              for alpha, beta in {z, +, -}
    fill!(y, 0)
    D = length(x)
    N = length(spins)

    parfor F = 0, ..., D-1
        decode target digits n_i = floor(F / M_i) mod d_i
        set target projections m_i = n_i - s_i
        acc = 0

        for i = 1, ..., N
            acc += h_z^(i) * m_i * x[F]

            if n_i > 0
                I  = F - M_i
                ms = (n_i - 1) - s_i
                acc += h_+^(i) * Lambda(+, s_i, ms) * x[I]
            end if

            if n_i < d_i - 1
                I  = F + M_i
                ms = (n_i + 1) - s_i
                acc += h_-^(i) * Lambda(-, s_i, ms) * x[I]
            end if
        end for

        for each pair term (i, j, J) in pair_terms
            for alpha in {z,+,-}
                for beta in {z,+,-}
                    ns_i = n_i - delta(alpha)
                    ns_j = n_j - delta(beta)

                    if 0 <= ns_i <= d_i - 1 and 0 <= ns_j <= d_j - 1
                        I   = F - delta(alpha) M_i - delta(beta) M_j
                        ms_i = ns_i - s_i
                        ms_j = ns_j - s_j
                        acc += J[alpha,beta]
                             * Lambda(alpha, s_i, ms_i)
                             * Lambda(beta, s_j, ms_j)
                             * x[I]
                    end if
                end for
            end for
        end for

        y[F] = acc
    end parfor

    return y
end function
\end{verbatim}
\end{widetext}
\onecolumngrid
\section{Higher-Power Spin Terms\label{appen:higher_powers}}
In the above text, we presented the matrix-free kernel for single-spin powers, which are easiest to describe, but the same algorithm applies to higher powers of local spin operators. Consider a local operator at site $i$ with matrix elements $\bra{n_i^{(F)}} \hat O_i \ket{n_i^{(I)}} \ne 0$, i.e.,  only when the target and source digits differ by a fixed integer shift $q$, such that $n_i^{(F)} = n_i^{(I)} + q$. In this case, a target digit $n_i$ can only receive a contribution from the source digit:
\begin{equation}
    n_i^{(\mathrm{src})}=n_i-q,
    \qquad
    I=F-qM_i,
    \label{eq:generic_local_shift}
\end{equation}
provided $0\le n_i-q\le d_i-1$.

For complex operators such as $(\Sp{i})^p (\Sz{i})^r (\Sm{i})^\ell$, the net shift is $q=p-\ell$. The total local coefficient is evaluated by applying the factors from right to left on the reconstructed source projection. Operators containing several possible shifts are first (analytically) decomposed into fixed-$q$ ladder monomials, and the resulting channels are summed. A particular application is the single-ion zero-field splitting (ZFS) Hamiltonian~\cite{atherton1993principles}:
\begin{equation}
    \Ham_{\mathrm{ZFS}}^{(i)}
    =D_i\left[(\Sz{i})^2-\frac{s_i(s_i+1)}{3}\right]
    +E_i\left[(\hat S_x^{(i)})^2-(\hat S_y^{(i)})^2\right].
    \label{eq:zfs_hamiltonian}
\end{equation}
Using the identity $(\hat S_x)^2-(\hat S_y)^2 = \frac{1}{2}(\hat S_+^2+\hat S_-^2)$, \eqref{eq:zfs_hamiltonian} decomposes into one diagonal channel ($q=0$) and two off-diagonal channels that shift the local digit by $q=\pm 2$. For a target projection $m_i = n_i - s_i$ (corresponding to target index $F$), the diagonal update is $y_F \mathrel{+}= D_i [m_i^2 - s_i(s_i+1)/3] x_F$. The off-diagonal updates utilize \eqref{eq:generic_local_shift} with $q=\pm 2$, where the two-step ladder coefficients are products of the elementary one-step coefficients:
\begin{align}
    C_+^{(2)}(s,m) = \Lambda(+;s,m+1)\Lambda(+;s,m), \quad
    C_-^{(2)}(s,m) = \Lambda(-;s,m-1)\Lambda(-;s,m).
\end{align}
For the $S_+^2$ channel, the admissibility condition is $n_i\ge 2$, the source index is
$I=F-2M_i$, and the coefficient is $C_+^{(2)}(s_i,m_i-2)$. For the $S_-^2$ channel, the
condition is $n_i\le d_i-3$, the source index is $I=F+2M_i$, and the coefficient is
$C_-^{(2)}(s_i,m_i+2)$. Thus the coefficient is always evaluated on the reconstructed source
projection, not on the target projection.
This logic handles single-ion anisotropy and quadrupolar terms (where one simply replaces the coefficients) with the same minimal integer arithmetic used for simple exchange terms.

\section{Spin--Boson Jaynes--Cummings Kernel\label{appen:spin_boson_jc}}
A truncated bosonic mode fits the same mixed-radix indexing once its occupation number is treated as a local digit $n_b \in \{0, 1, \dots, n_{\max}\}$ with multiplier $M_b$. For clarity, consider spin-$1/2$ sites with digits $\sigma_i=0$ for spin down and $\sigma_i=1$ for spin up. A single spin gives the Jaynes--Cummings model~\cite{JC_Model, JC_model_tutt}, while several spins coupled to the same truncated mode give the Tavis--Cummings form~\cite{TC_model}:
\begin{equation}
    \Ham_{\mathrm{JC}}
    =\omega_b\hat a^\dagger\hat a
    +\sum_i \Omega_i \Sz{i}
    +\sum_i\left(g_i\hat a^\dagger\Sm{i}+g_i^*\hat a\Sp{i}\right).
    \label{eq:jc_hamiltonian}
\end{equation}
For a target state with boson occupation $n_b$ and spin digits $\{\sigma_i\}$, the diagonal contribution is
\begin{equation}
    y_F \mathrel{+}=
    \left(\omega_b n_b+\sum_i \Omega_i(\sigma_i-\tfrac{1}{2})\right)x_F.
\end{equation}
The interaction terms involve simultaneous shifts of the boson digit and one spin digit. The two
rotating-wave channels are:
\begin{enumerate}
    \item $\hat a^\dagger\Sm{i}$ contributes
    only when $n_b>0$ and $\sigma_i=0$. The source has occupation $n_b-1$ and spin up, so
    \begin{equation}
        I=F-M_b+M_i,
        \qquad
        y_F\mathrel{+}=g_i\sqrt{n_b}\,x_I.
    \end{equation}
    \item $\hat a\Sp{i}$ reaches a target with one fewer boson and spin up. It contributes only
    when $n_b<n_{\max}$ and $\sigma_i=1$. The source has occupation $n_b+1$ and spin down, so
    \begin{equation}
        I=F+M_b-M_i,
        \qquad
        y_F\mathrel{+}=g_i^*\sqrt{n_b+1}\,x_I.
    \end{equation}
\end{enumerate}
The square-root factor is the bosonic matrix element evaluated on the reconstructed source state.

While this example assumes spin-$1/2$ for simplicity, the construction generalizes to higher spins or multiple bosonic modes. A term $g_{\mu i}\hat a_\mu^\dagger\Sm{i}$ for mode $\mu$ and spin $i$ uses the offset $I = F - M_\mu + M_i$ and includes the standard spin-ladder matrix element in the coefficient. Counter-rotating terms are handled identically. This demonstrates that arbitrary mixed spin-boson systems can be propagated matrix-free using the same uniform logic of integer offsets and local matrix elements.
\section{Step-by-step illustration of the lookup mechanism\label{sec:lookup_mech}}
To present the abstract lookup structure more concretely, let's consider a minimal toy example: a one-dimensional chain of $N=4$ spin-$1/2$ particles restricted to the zero-magnetization sector ($M=0$, half-filling). The full Hilbert space has a dimension of $D = 2^4 = 16$. However, the constraint $M=0$ restricts the size of the admissible states. The restricted basis dimension is therefore $N_{\mathrm{res}} = \binom{4}{2} = 6$.

We map the local states to binary digits ($\uparrow \equiv 1$, $\downarrow \equiv 0$). The six admissible tuples $\boldsymbol{\lambda}$ and their dense-packed-storage indices $I_{\mathrm{packed}} \in \{0, \dots, 5\}$ are shown in Table \ref{tab:toy_model}.
\begin{table}[h]
    \centering
    \begin{tabular}{c c}
        \hline\hline
        Physical State $\boldsymbol{\lambda}$ & Storage Index $I_{\mathrm{packed}}$ \\
        \hline
        $|0011\rangle$ & 0 \\
        $|0101\rangle$ & 1 \\
        $|0110\rangle$ & 2 \\
        $|1001\rangle$ & 3 \\
        $|1010\rangle$ & 4 \\
        $|1100\rangle$ & 5 \\
        \hline\hline
    \end{tabular}
    \caption{State mappings for an $N=4$, $M=0$ restricted basis under zero magnetization constraint.}
    \label{tab:toy_model}
\end{table}

Now, suppose we need to evaluate the action of the first half of an off-diagonal flip-flop operator, $H_{\mathrm{off}} = \frac{J}{2} S_2^+ S_0^-$, on the target state with packed index $F_{\mathrm{tgt}} = 5$, which corresponds to $\ket{1100}$. (This is one term of the Hermitian pair $\frac{J}{2}(S_2^+ S_0^- + S_2^- S_0^+)$; the conjugate term contributes when the algorithm reaches the partner target $\ket{1001}$, where the inverse local move yields the source $\ket{1100}$.)

The algorithm proceeds as follows:
\begin{enumerate}
    \item \textbf{State Unpacking:} We first read the target index $F_{\mathrm{tgt}} = 5$ and fetch the corresponding physical state (or corresponding labels) $\ket{1100}$
    \item \textbf{Local Operator Action:} We then apply the physical rules of $H_{\mathrm{off}}$ to the bit-string $\ket{1100}$, in our case the operator $S_2^+ S_0^-$ annihilates a spin at site 0 and creates one at site 2. So the source corresponding to the action operator $S_2^+ S_0^-$ on state $\ket{1100}$ is $\ket{1001}$.
\begin{equation}
        \ket{1100} \xleftarrow{S_2^+ S_0^-} \ket{1001}.
\end{equation}
The candidate source label $\boldsymbol{\lambda}^{(\mathrm{src})}$ is now identified as $1001$~\footnote{We assume the source label generation is efficient based on the physics of the system}.
    \item \textbf{The Membership Lookup:} Now one calls for the lookup function on the source labels:
    \begin{equation}
        I_{\mathrm{src}} = \mathrm{lookup}(1001).
    \end{equation}
    Because $1001$ is a valid label in our restricted basis, the lookup succeeds and returns $I_{\mathrm{src}} = 3$. If the operator had generated a source-state outside the sector (e.g., $|1111\rangle$), the lookup would flag the state as invalid, and the matrix element would be evaluated as zero.
    \item \textbf{Vector Accumulation:} Having successfully identified both indices, the algorithm updates the target amplitude in the output vector:
    \begin{equation}
        y[5] \mathrel{{+}{=}} \left(\frac{J}{2}\right) x[3].
    \end{equation}
\end{enumerate}
\end{document}